\documentclass[5p,twocolumn,times]{elsarticle}
\usepackage{lipsum}
\usepackage{lineno,hyperref}
\modulolinenumbers[5]
\usepackage[pdftex]{color}
\usepackage[font=footnotesize,labelfont=bf]{caption}
\usepackage[font=footnotesize,labelfont=bf]{subcaption}
\journal{Journal of \LaTeX\ Templates}
\usepackage{colortbl}
\usepackage{xcolor}

\usepackage{tikz-cd}
\usepackage{amssymb}

\biboptions{compress}

\usepackage[figuresright]{rotating}

\begin{document}

\begin{frontmatter}


\title{Impact of reproduction-mobility trade-off on biodiversity in rock-paper-scissors models in changing environmental conditions}


\address[1]{Institute for Biodiversity and Ecosystem
Dynamics, University of Amsterdam, Science Park 904, 1098 XH
Amsterdam, The Netherlands}
\address[2]{School of Science and Technology, Federal University of Rio Grande do Norte\\
Caixa Postal 1524, 59072-970, Natal, RN, Brazil}
\address[3]{ Edmond and Lily Safra International Institute of Neuroscience, Santos Dumont Institute,
Av Santos Dumont 1560, 59280-000, Macaiba, RN, Brazil}
\address[4]{Department of Computer Engineering and Automation, Federal University of Rio Grande do Norte, Av. Senador Salgado Filho 300, Natal, 59078-970, Brazil}

\author[1,2]{J. Menezes}  
\author[3,4]{E. Rangel} 

\begin{abstract}
We investigate a tritrophic system in which organisms' energy depletion, resulting from failed selection attempts, leads to a partial loss of capacity to win the cyclic spatial game.
The energy required to maintain optimal organism fitness may be impacted by changes in environmental conditions, increasing the death risk due to accelerated deterioration of health.
We simulate the evolutionary behavioural strategy performed by individuals of one species, which consists in balancing efforts dedicated to reproduction and mobility to minimise the chances of death by lack of energy. We show that the unevenness introduced by the trade-off strategy unbalances the rock-paper-scissors model, with the predominant species profiting from enemies' lower birth rate of enemies. Quantifying the spatial patterns, we demonstrate that the characteristic length scale of single-species domains decreases as energy loss accelerates due to environmental changes. The erosion in the spatial patterns provoked by the reproduction-mobility trade-off benefits biodiversity, with coexistence probability rising for faster energy depletion and higher trade-off factors. The findings have implications for ecologists seeking to understand the impact of survival behaviour on biodiversity promotion.
\end{abstract}


\end{frontmatter}



\section{Introduction}

In ecology, much attention has been devoted to investigating the role of organisms' behaviour in ecosystem formation and stability \cite{ecology,climatechange,adap2}. Plenty of evidence shows that, in response to environmental changes, animals perform behavioural strategies, adapting their movement following signals received from the neighbourhood \cite{adaptive1,adaptive2,Dispersal,BENHAMOU1989375,coping}. 
Many researchers have reported that recognising hostile regions or lack of natural resources is an evolutionary ability that allows organisms to flee from enemies, thus prolonging their survival in adverse environmental changes
\cite{foraging,BUCHHOLZ2007401,adap-self,bac-co,bac-co2,Causes,howdirectional,Agg,combination,adaptive-epl,adaptive-jc,locally,eloi}. This accurate adaptability observed in nature has inspired engineers to create new generations of robots whose movement imitates the animals' locally adaptive strategies \cite{animats}. Software engineers have also used this to build artificial intelligence-based algorithms following the 
animals' self-adaptive foraging behaviour, thus optimising computer systems to respond to changing conditions \cite{fora-adap-book}.

We investigate the spatial rock-paper-scissors model, considering that individuals may face energy depletion when failing to defeat enemies in the spatial game \cite{Reichenbach-N-448-1046,Coli,bacteria,Allelopathy,starvation,reversal}. 
In this letter, we assume that changes in environmental conditions impact the energy needed to maintain an organism's optimal fitness, leading to an accelerated decline in its health. Recently, it has been shown that, in cyclic games,
organisms that strategically redirect energy from reproduction activity to increase the dispersal rate have more chances of recovering energy, thus minimising the risk of dying because of energy depletion \cite{primeiro,tradeoff0,tradeoff2,tradeoff3}. 
Moreover, controlling the birth rate to redirect energy to explore more extensive areas reduces the average probability of individuals being caught by enemies, besides being less likely to die from energy-loss-related issues \cite{primeiro}. The consequence is the prolongation in the expected survival time, proving the profitability of the self-adaptive strategy in extended longevity. 

This raises further questions about the impact on the equilibrium in the cyclic spatial game and the consequent effects on biodiversity maintenance \cite{uneven,unevenmob,Adap,AdapII,Moura,multiple}.
Because of this, in this work, we aim to discover: i) how is the rock-paper-scissors game unbalanced by the trade-off strategy of one out of the species;
ii) which species predominates in the spatial game if the trade-off strategy is executed?;
iii) what is the dependence of the average size of the single-species spatial domains with the trade-off factor?; 
iv) how does the speed of the organisms' energy depletion impact the spatial pattern transformation?; v) which species predominates in the spatial game when the trade-off strategy is performed?; vi) is biodiversity promoted or jeopardised by the trade-off strategy?

\begin{figure}\centering
\includegraphics[width=45mm]{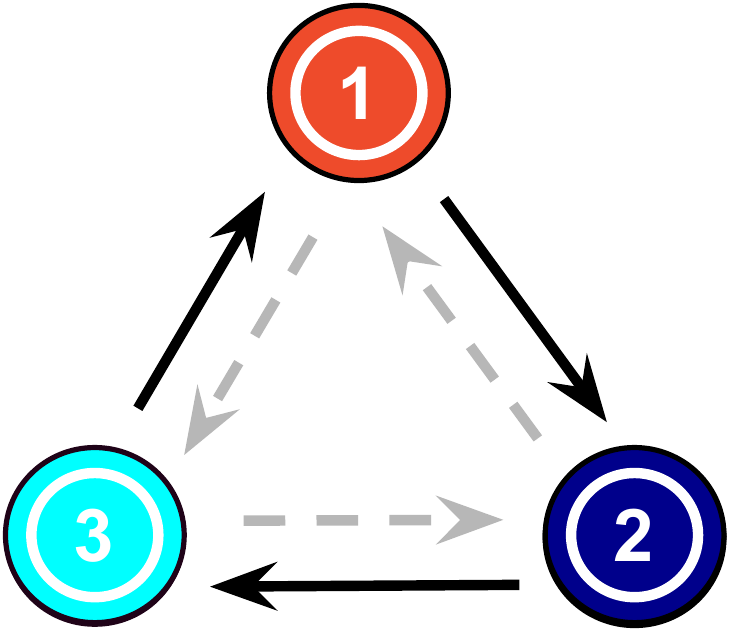}
\caption{Illustration of selection rules in our spatial rock-paper-scissors game. Solid black arrows indicate the cyclic dominance of individuals of species $i$ over organisms of species $i+1$; the dashed grey arrows show that reversal selection interactions may occur when individuals of species $i$ are weak.}
	\label{fig1}
\end{figure}
\begin{figure*}[t]
\centering
        \begin{subfigure}{.23\textwidth}
        \centering
        \includegraphics[width=40mm]{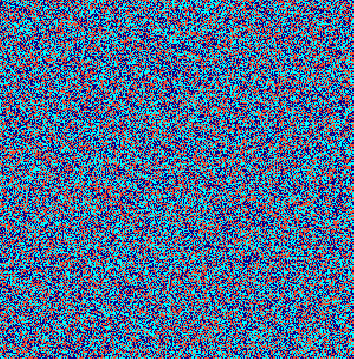}
        \caption{}\label{fig2a}
    \end{subfigure} %
    \begin{subfigure}{.23\textwidth}
        \centering
        \includegraphics[width=40mm]{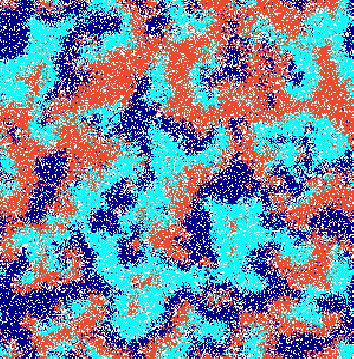}
        \caption{}\label{fig2b}
    \end{subfigure} %
       \begin{subfigure}{.23\textwidth}
        \centering
        \includegraphics[width=40mm]{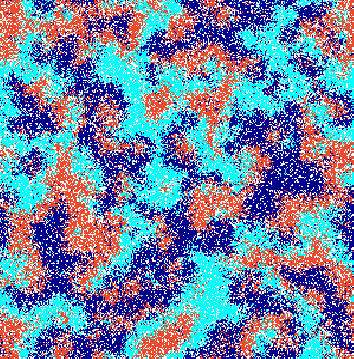}
        \caption{}\label{fig2c}
    \end{subfigure} %
           \begin{subfigure}{.23\textwidth}
        \centering
        \includegraphics[width=40mm]{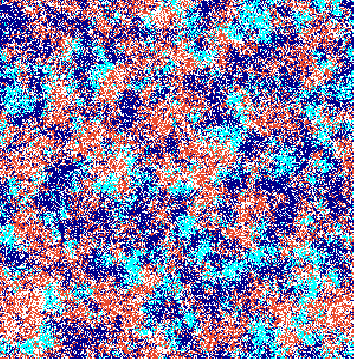}
        \caption{}\label{fig2d}
    \end{subfigure} %
\caption{Snapshots from simulations of the rock-paper-scissors game with various trade-off factors. The realisations ran in lattice with $300^2$ grid points, starting from the initial conditions in Fig.~\ref{fig2a}. The organisms' spatial configuration after $3000$ generations is shown in Figs.~\ref{fig2b}, ~\ref{fig2c}, and ~\ref{fig2d}, for Simulation A ($\beta=0.0$), B ($\beta=0.5$), and C ($\beta=1.0$), respectively. The colours follow the scheme in Fig. \ref{fig1}, with red, dark blue, and cyan depicting individuals of species $1$, $2$, and $3$, respectively. White dots show empty sites. 
The dynamics of the organisms' spatial organisation during the entire simulations A, B, and C are shown in Videos https://youtu.be/wCxQi9ipk0U, https://youtu.be/zKL04zY8nyM, and https://youtu.be/5z4Do5L8EiY.}
  \label{fig2}
\end{figure*}
\begin{figure}
    \centering
        \begin{subfigure}{.48\textwidth}
        \centering
        \includegraphics[width=85mm]{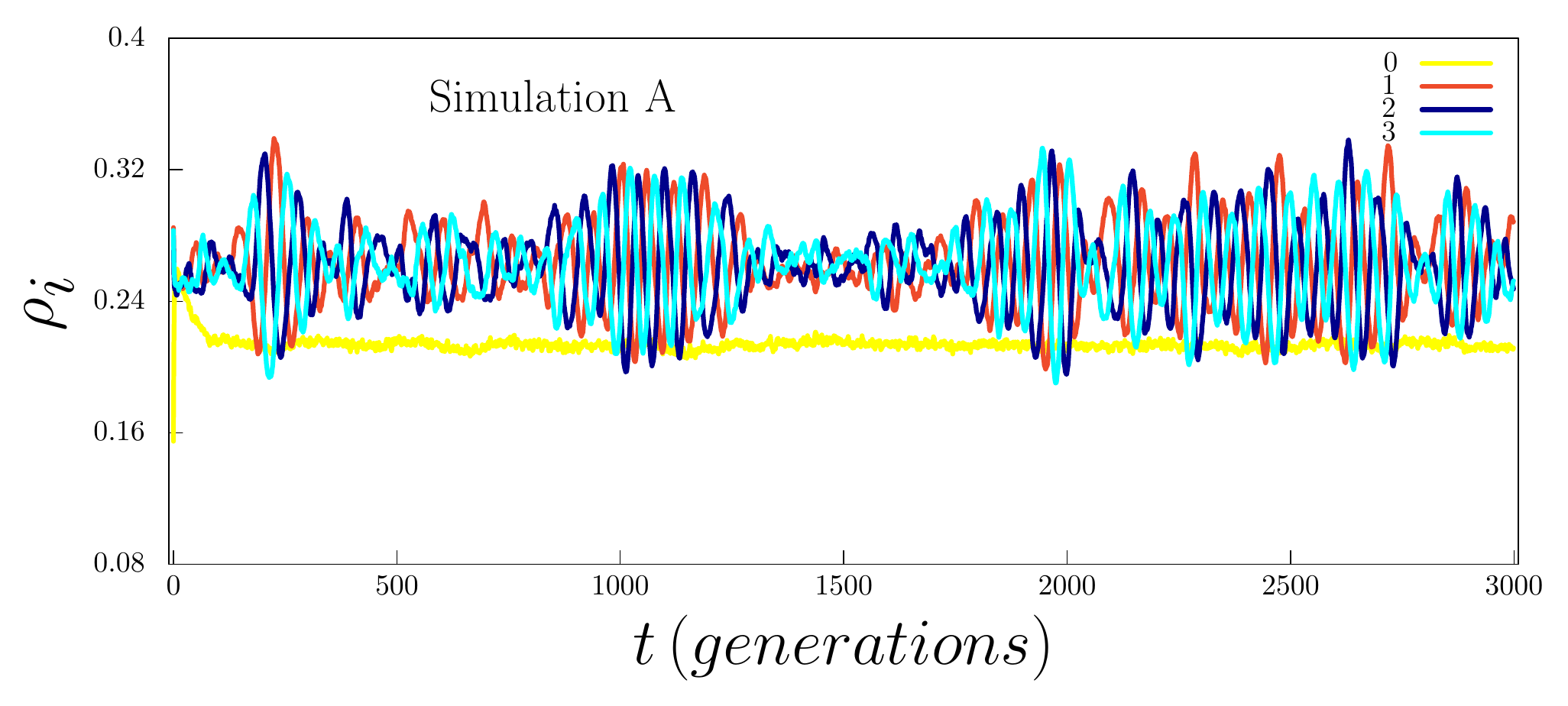}
        \caption{}\label{fig3a}
    \end{subfigure}\\
       \begin{subfigure}{.48\textwidth}
        \centering
        \includegraphics[width=85mm]{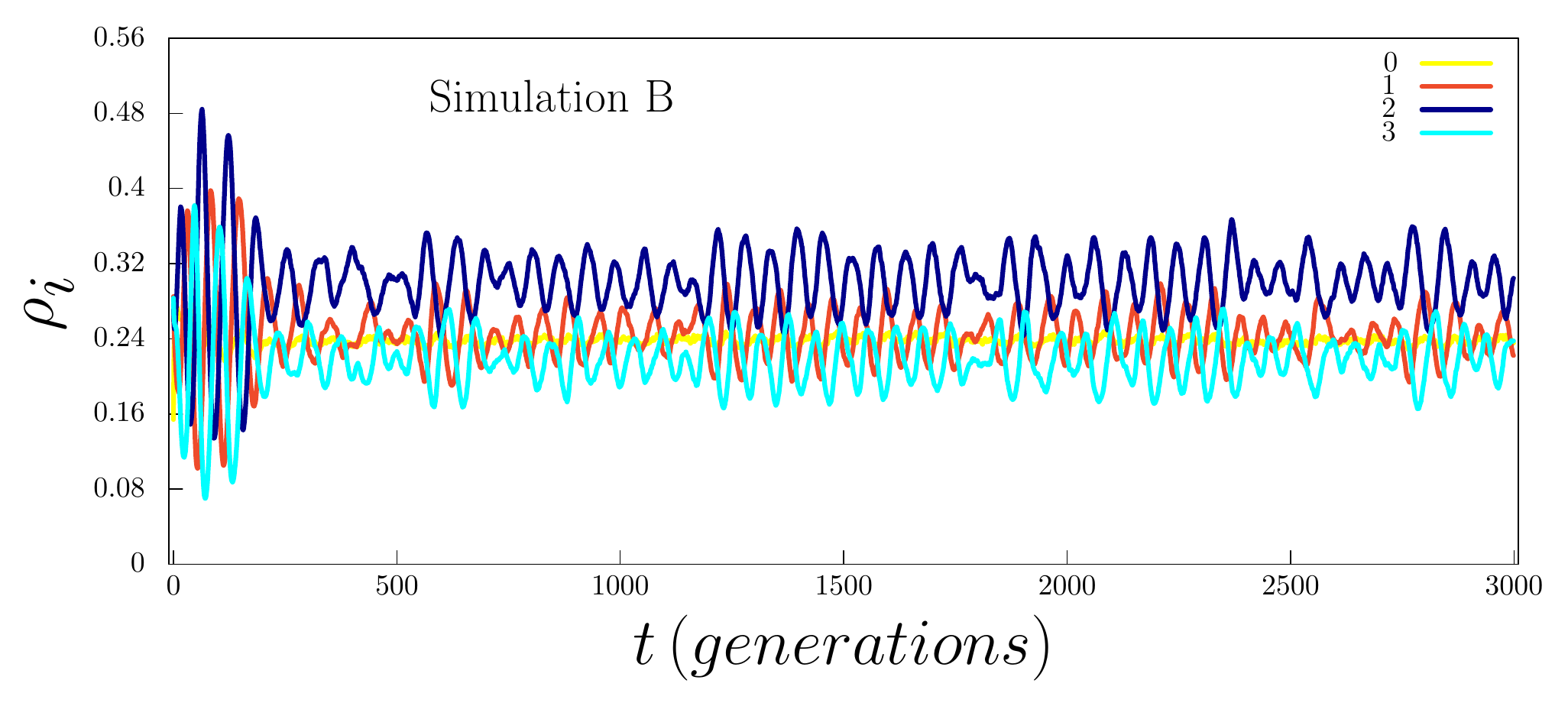}
        \caption{}\label{fig3b}
        \end{subfigure}\\
       \begin{subfigure}{.48\textwidth}
        \centering
        \includegraphics[width=85mm]{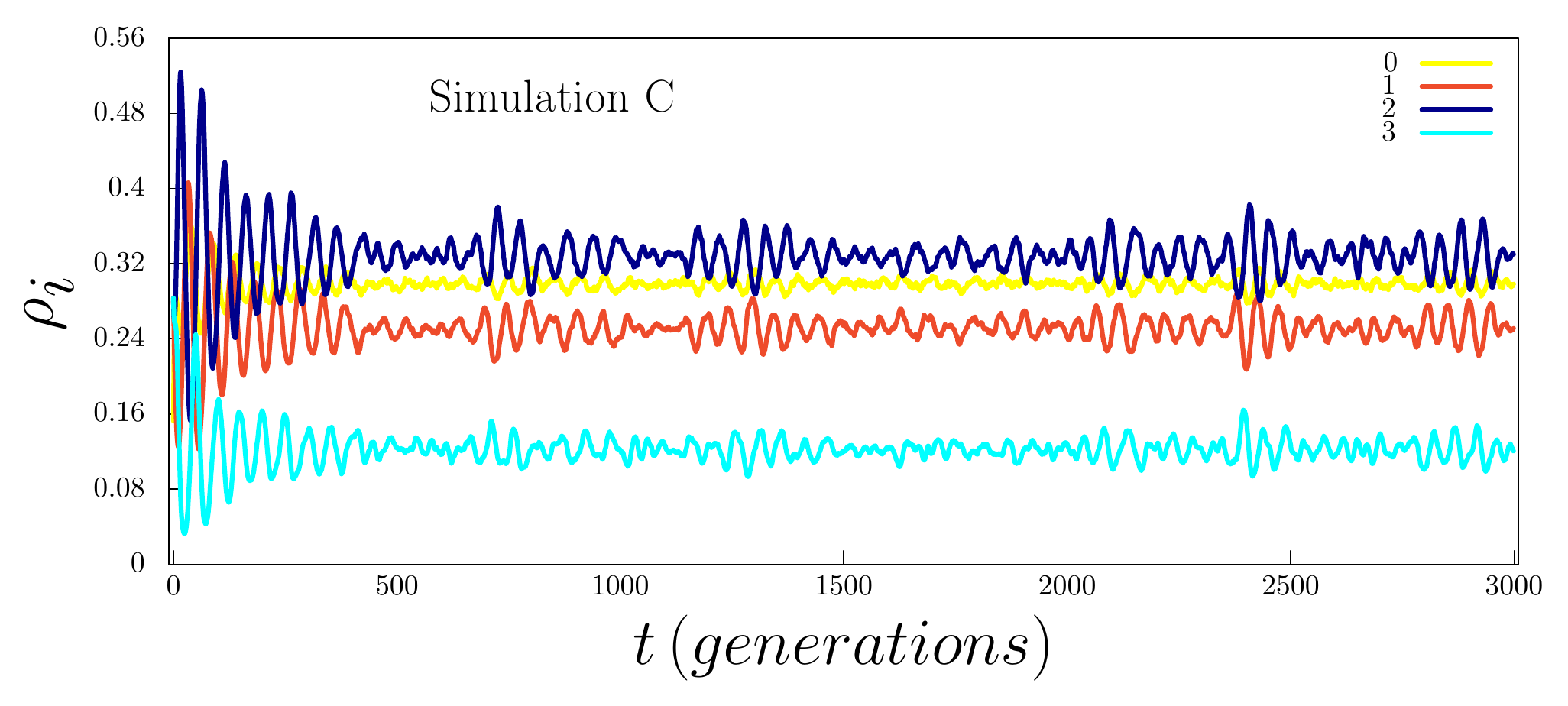}
        \caption{}\label{fig3c}
    \end{subfigure}
\caption{Temporal dependence of the species densities. Figures \ref{fig3a}, \ref{fig3b}, and \ref{fig3c} shows the population dynamics 
in simulations A (https://youtu.be/wCxQi9ipk0U	), B (https://youtu.be/zKL04zY8nyM), and C (https://youtu.be/5z4Do5L8EiY), respectively.
The colours follow the scheme in Fig.~\ref{fig1}; yellow lines show the density of empty spaces.
}
\label{fig3}
\end{figure}
\section{The Model}

We model a cyclic out-competition of organisms of three species, following the popular rock-paper-scissors game rules - scissors cut paper, paper wraps rock, rock crushes scissors. 
Figure \ref{fig1} illustrates the spatial game with black arrows indicating that individuals of species $i$ eliminate organisms of species $i+1$, with $i=1,2,3$ - we assume the cyclic identification $i=i+3\,\alpha$, where $\alpha$ is an integer. 

Following the model introduced in Ref.~ \cite{primeiro}, we consider that the organism's strength depends on the success of its selection attempts. This means that if an organism does not manage to eliminate a neighbouring individual to control the local natural resources, its energy drops. Once weak, the individual's selection capacity diminishes, thus becoming possible to be killed by a reversal selection interaction, as illustrated by the dashed grey arrows in Fig.~\ref{fig1}.

We work with a three-state energy configuration: high, intermediate, and low, named levels $3$, $2$, and $1$, respectively. Accordingly, whenever an individual successfully eliminates a neighbour, the energy level may transition to a higher state or maintain at the highest level. Otherwise, 
the energy may decay, or a low-energy organism may die.
Moreover, the organism's capacity to win the spatial game depends on its energy level. Hence, only a high-energy individual of species $i$ has $100\%$ chances of killing an individual of species $i+1$, which does not depends on the energy level of the eliminated organism. But, intermediate-energy and low-energy individuals partially lose selection capacity, compromising the spatial game's cyclic advantage.

As a responsive behavioural strategy, weak organisms of species $1$ perform a reproduction-mobility trade-off to explore a more significant fraction of the territory, thus
increasing the chances of selecting vulnerable organisms and accelerating energy recovery. The proportion of energy redirected from producing offspring to grow dispersal rate is controlled by the real parameter $\beta$, with $0\leq \beta \leq 1$, the trade-off factor.

\subsection{Stochastic simulations}

We use square lattices with $\mathcal{N}$ grid sites and periodic boundary conditions to perform our stochastic simulations. Our implementation follows the May-Leonard models, where each grid site either contains one individual or remains empty \cite{leonard}. Using the notation in Fig.~\ref{fig1}, we define the total number of organisms of species $i$ in the lattice as $I_i$, with $i=1,2,3$. Therefore the density of organisms of species $i$ is defined as $\rho_i = I_i/\mathcal{N}$, with $i=1,2,3$; complementarily, the density of empty spaces is $\rho_0 = 1 -\rho_1-\rho_2-\rho_3$.

Initially, each organism is allocated to a random grid site. The initial conditions are built with $\rho_1=\rho_2=\rho_3$. It is considered the maximum number of organisms that fit in the lattice, which constrains $\rho_0$ being minimum at the initial state. The algorithm uses the Moore neighbourhood to execute the stochastic simulations: each organism may interact with one of its eight immediate neighbours. The possible interactions are described as follows:
\begin{itemize}
\item
Selection: an individual of species $i$ kills a neighbouring individual of species $i+1$ (direct selection) or $i-1$ (reversal selection), leaving the grid site vacant. 
The probability $s$ defines the chances of a selection interaction being stochastically chosen, whose implementation depends on the organisms' energy state:

\par
\begin{tikzcd}
\centering
i_l\,\,\,\, (i+1)_{k} \rar{s}\,\, &\,\, i_l \,\,\,\,\, \otimes, \,\,\, l \geq k\,\,\,\,\,\,\,\\
i_l\,\,\, (i+1)_{k} \rar{\eta\,\gamma s/2}\,\, &\,\, i_l \,\,\, \otimes, \,\, k-l=\eta\\
i_l\,\,\, (i-1)_{k} \rar{\eta\,\gamma s/2}\,\, &\,\, i_l \,\,\, \otimes, \,\, l-k=\eta
\end{tikzcd}

where $\eta$ is the difference between the energy states between the active and passive individuals; 
$\gamma$ is a real parameter, with $0\leq \gamma \leq 1$, representing the probability of the individual's energy state allowing reversal selection interaction. Additionally, we define 
the energy depletion factor $\varepsilon$, a real parameter, with $0\leq \varepsilon \leq 1$, representing the probability of the active individual transitioning to a lower energy state or 
a low-energy individual dying in case of the selection interaction not to be implemented.

\item
Reproduction: an empty space is filled by a new high-energy organism of species $i$. Although this action is chosen with 
probability $r$, because of the reproduction-mobility trade-off, 
intermediate and low-energy organisms of species $1$ 
purposely reduces the chances of generating offspring to
$(1-\beta)\,r$:
\par
\begin{tikzcd}
\centering
1_l\,\,\,\, \otimes \rar{r}\,\, &\,\, 1_l \,\,\,\,\, 1_l, \,\,\,\,\,\,\,\,\,l=3\\
1_l\,\,\, \otimes \rar{(1-\beta)\,r}\,\, &\,\, 1_l \,\,\, 1_l, \,\,\,\,\, l < 3\\
i_l\,\,\,\, \otimes \rar{r}\,\, &\,\, i_l \,\,\,\,\, i_l, \,\,\,\,\,\,\,\,\, i=2,3
\end{tikzcd}

where $\otimes$ means an empty space and $l$ is the individual energy state.
\item
Mobility: an individual switches grid site with either another organism or an empty space. In general, individuals move with probability $m$, but due to the trade-off between reproduction and mobility, intermediate and low-energy organisms of species $1$ move faster than the others, with probability
$m+\beta\,r$:

\par
\begin{tikzcd}
\centering
1_l\,\,\,\, \odot \rar{m}\,\, &\,\, \odot \,\,\,\,\, 1_l, \,\,\,\,\,\,\,\,\,l=3\\
1_l\,\,\, \odot \rar{m\,+\beta\,r}\,\, &\,\, \odot \,\,\, 1_l, \,\,\,\,\, l < 3\\
i_l\,\,\,\, \odot \rar{m}\,\, &\,\, \odot\,\,\,\,\, i_l, \,\,\,\,\,\,\,\,\, i=2,3
\end{tikzcd}

where $\odot$ means an empty space or an individual of any species.
\end{itemize}

The stochasticity of the interaction implementation follows the steps:
\begin{enumerate} 
\item
an active individual of any species is randomly picked among all organisms in the lattice;
\item
one of the interactions is randomly chosen according to a set of fixed probabilities;
\item
one of the eight immediate neighbours is randomly drawn to suffer the chosen interaction (selection, reproduction, or mobility).
\end{enumerate} 
If not stated otherwise, we obtained the results presented in this work using the interaction probabilities: $s = r = m = 1/3$, and $\gamma=0.5$. However, we have performed simulations with different parameters to ensure that our main conclusions are independent of the stochasticity parameters.
Each interaction implementation represents a single time step. Thus, we define our time unit as one generation, which is completed when $\mathcal{N}$ time steps occur.


\section{Organisms' spatial organisation}

The initial step of our research is to observe the changes in the spatial patterns produced by the trade-off strategy of individuals of species $1$. To achieve this objective, three simulations were conducted in lattices with $300^2$ grid sites, beginning from the random initial condition shown in Fig.~\ref{fig2a}:
\begin{itemize}
\item
Simulation A: $\beta=0.0$ - organisms of species $1$ do not allocate energy towards foraging, even when in a low or intermediate energy state;
\item
Simulation B: $\beta=0.5$ - weak individuals of species $1$ cease $50\%$ reproduction to enhance their mobility rates;
\item
Simulation C: $\beta=1.0$ - intermediate and low-energy organisms of species $1$ redirect $100\%$ energy from reproduction to mobility.
\end{itemize}

We captured snapshots after $3000$ generations of 
each realisation.
The final spatial individuals' organisation is shown in Figs. \ref{fig2b}, \ref{fig2c}, and \ref{fig2d}, for 
Simulations A, B, and C, respectively.
Additionally, videos https://youtu.be/wCxQi9ipk0U, https://youtu.be/zKL04zY8nyM, and https://youtu.be/5z4Do5L8EiY show how the spatial configuration changes during Simulations A, B, and C.
We utilise the same colour scheme of Fig.~\ref{fig1} to depict individuals: red, dark blue, and cyan dots represent organisms of species $1$, $2$, and $3$, respectively. White dots show empty spaces. 

The outcomes of Simulation A show what happens in a balanced scenario where all organisms of every species face energy depletion without a responsive trade-off strategy. Due to the cyclic selection dominance, organisms of the same species segregate into departed spatial domains, as seen in Fig.~\ref{fig2b}. Because of the symmetry of the rock-paper-scissors game rules, the average size of red, dark blue, and cyan regions are the same. Moreover, the death of low-energy organisms results in many empty spaces within the single-species domains. 

However, if half energy usually spent with reproduction is redirected to accelerate the movement of weak individuals of species $1$, the spatial patterns are significantly altered, as shown in Fig.~\ref{fig2c}. Accordingly, in Simulation B, the average concentration of empty spaces is higher within orange areas than dark blue and cyan domains. This happens because of the reduced average reproduction rate of individuals of species $1$. This effect is further emphasised in Simulation C, where weak individuals of species $1$ stop reproducing to move more quickly, as shown in the right panel of Fig.~\ref{fig2d}.
Comparing the snapshots of Simulations B and C, we conclude that the average size of the single-species domains decreases as $\beta$ grows. Moreover, as intermediate and low-energy organisms of species $1$ move faster than the others, they explore larger areas. Because of this, the average size of the orange areas is the largest, as seen in Figs.~ \ref{fig2c} and ~ \ref{fig2d}, for Simulations B and C. Also, as the trade-off factor grows, the size of the areas occupied by species 3 (represented in cyan) is notably reduced.

\begin{figure}
\centering
    \begin{subfigure}{.49\textwidth}
        \centering
        \includegraphics[width=85mm]{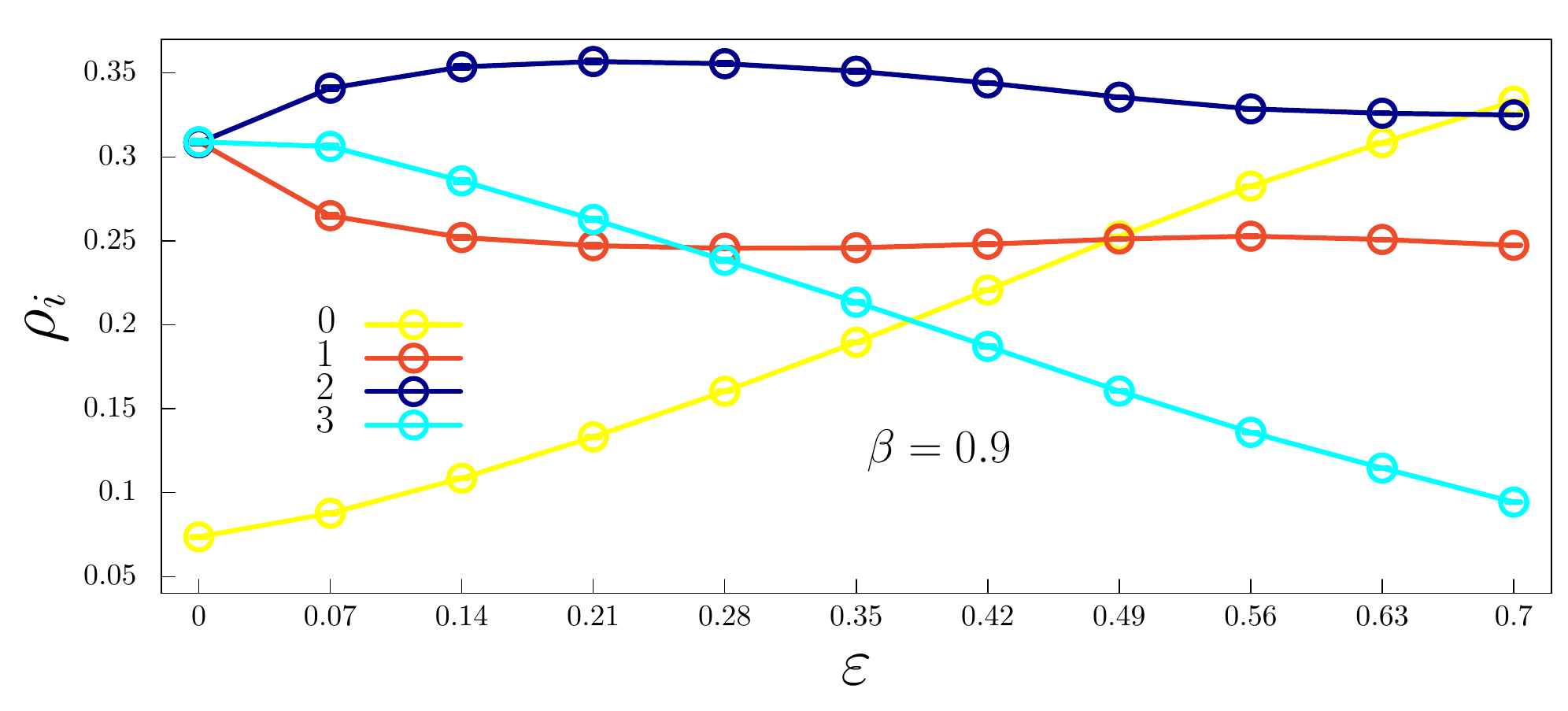}
        \caption{}\label{fig4a}
    \end{subfigure} %
       \begin{subfigure}{.49\textwidth}
        \centering
        \includegraphics[width=85mm]{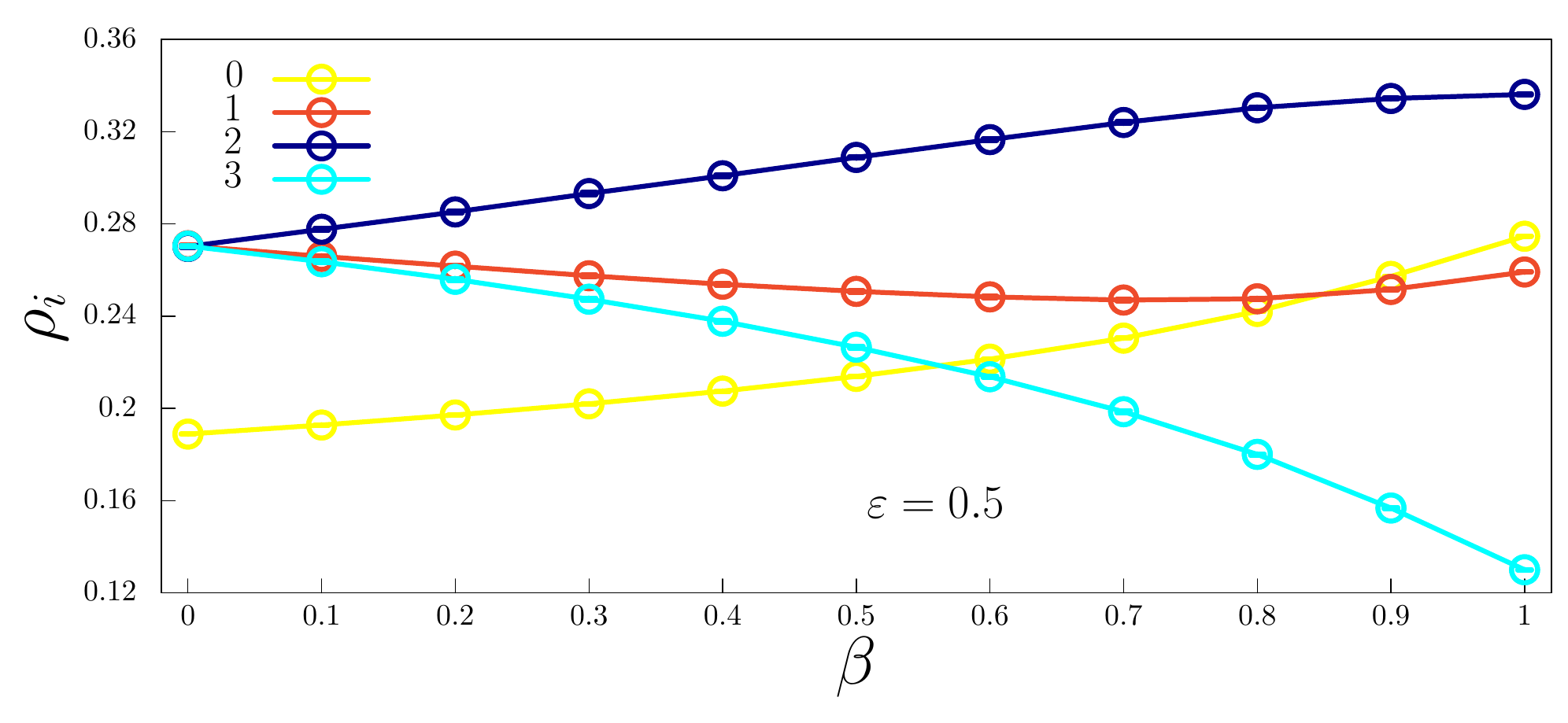}
        \caption{}\label{fig4b}
    \end{subfigure} %
    \caption{Species densities in terms of the model parameters. Figures \ref{fig4a} and \ref{fig4b} depict the dependence of the species densities on the energy depletion and trade-off factors, respectively. The outcomes were computed using sets of $100$ simulations; the error bars indicate the standard deviation. The colours follow the scheme in Fig.~\ref{fig1}; yellow lines show the density of empty spaces.}
  \label{fig4}
\end{figure}
\begin{figure}[t]
\centering
    \begin{subfigure}{.49\textwidth}
        \centering
        \includegraphics[width=85mm]{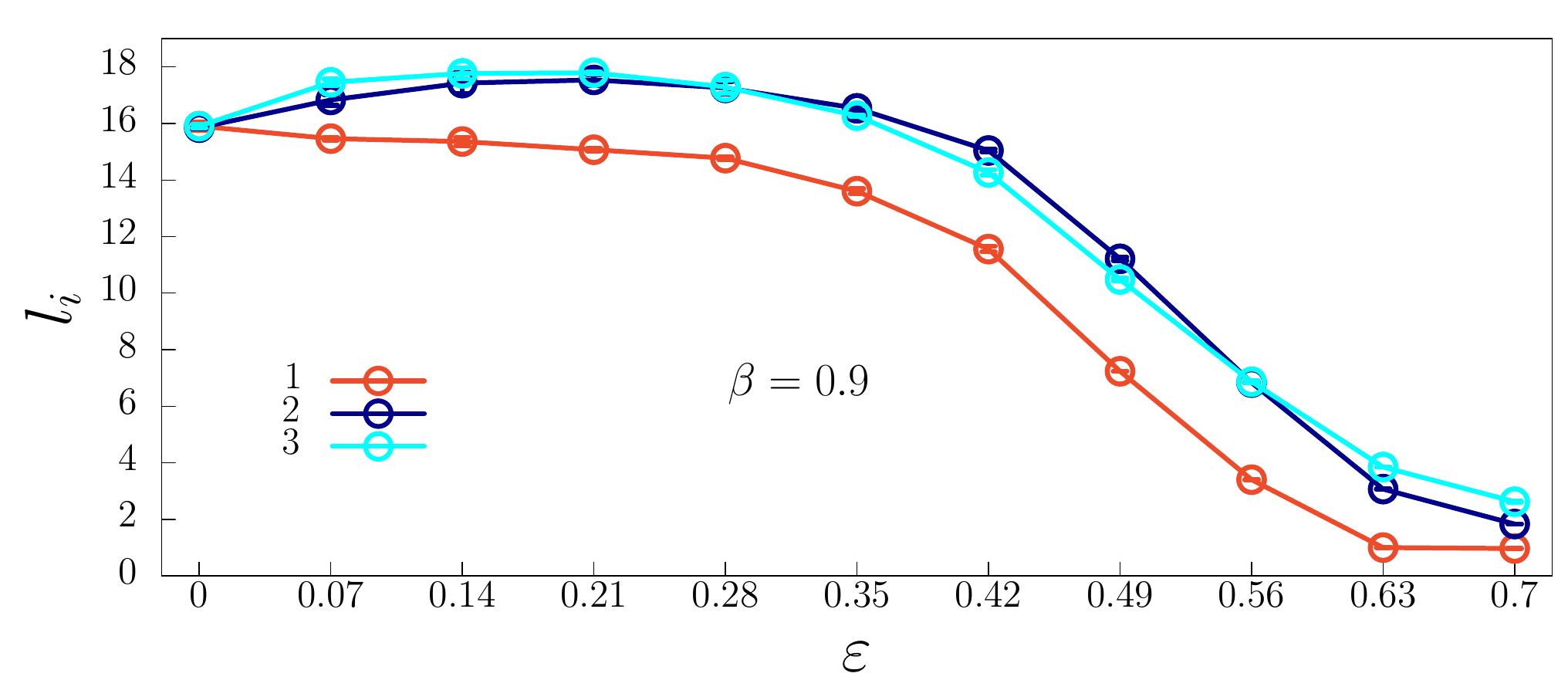}
        \caption{}\label{fig5a}
    \end{subfigure} %
       \begin{subfigure}{.49\textwidth}
        \centering
        \includegraphics[width=85mm]{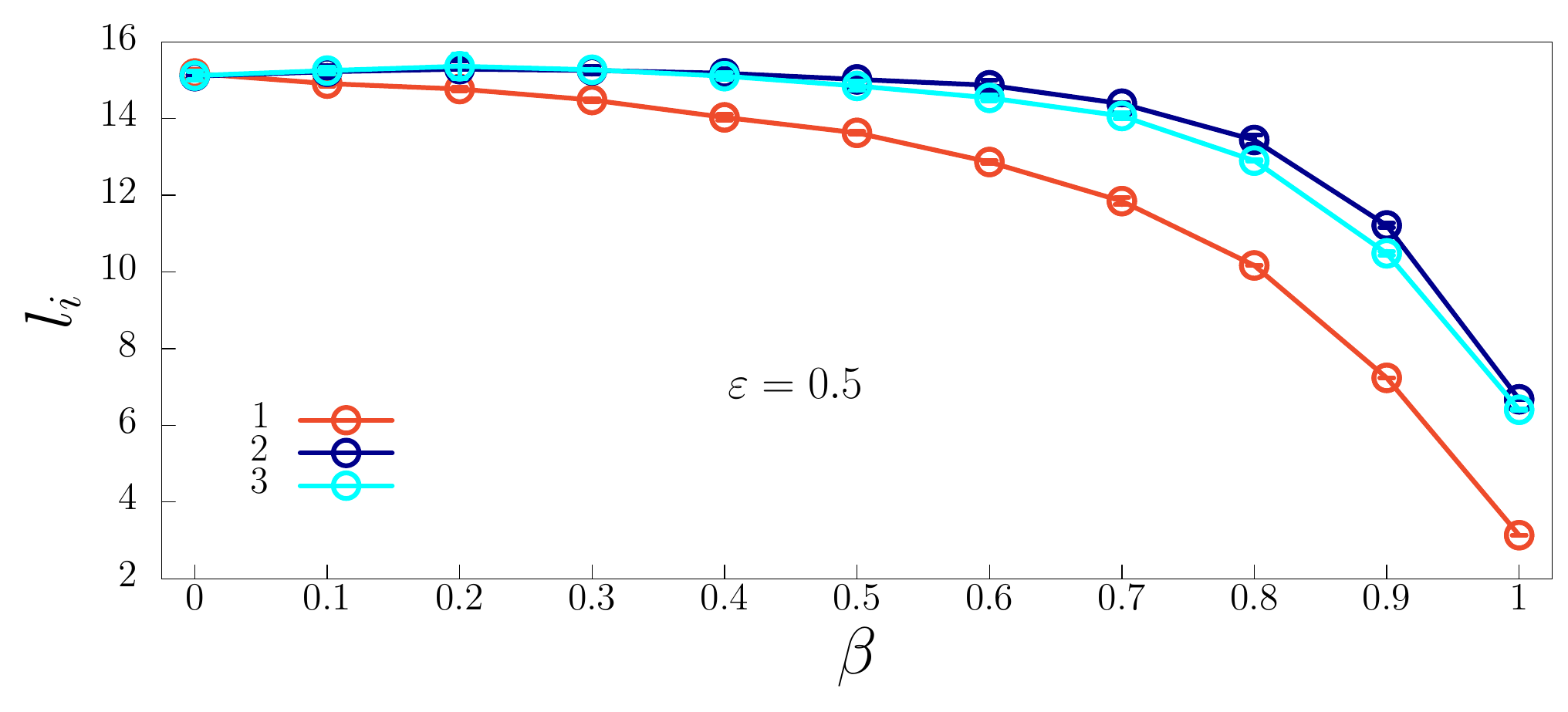}
        \caption{}\label{fig5b}
    \end{subfigure} 
    \caption{The characteristic length scale of single-species spatial domains in terms the model parameters. Figures \ref{fig5a} and \ref{fig5b} shows how the energy depletion and trade-off factors interferes with spatial patters, respectively.
The colours follow the scheme in Fig.~\ref{fig1}; the error bars show the standard deviation from sets of $100$ simulations with different random initial conditions.}
  \label{fig5}
\end{figure}

\section{Species Densities}
\label{sec5}
We now focus on the effects of the trade-off between reproduction and mobility on population dynamics. 
For this purpose, we first observe the temporal 
dependence of the species densities in simulations $A$, $B$, and $C$, which appear in Figs.~\ref{fig3a}, ~\ref{fig3b}, and ~\ref{fig3c}, respectively. The colours follow the scheme in Fig.~\ref{fig1}; yellow lines represent the density of empty spaces.

Initially, a high fluctuation in species densities is observed in the transient pattern formation period. 
While organisms of species $1$ do not employ the trade-off survival strategy, the spatial game is in equilibrium, with species cyclically alternating the territory dominance with the same average species density, as shown in Fig.~\ref{fig3a} - the oscillations are inherent to the cyclic dominance of the rock-paper-scissors game.
However, the population dynamics are impacted if individuals of species $1$ behave differently from those of species $2$ and $3$. Figures \ref{fig3b} and \ref{fig3c} show that species $2$ becomes predominant in the spatial game, with the population of species $3$ strongly declining. Also, we observe that the average density of empty spaces is higher in Fig.~\ref{fig3c}.

\subsection{The influence of the depletion energy and trade-off factors}

We quantify the dependence of average species densities on the model parameters by running two groups of simulations: i) for a fixed $\beta=0.9$, we assume that the energy depletion factor varies in the range $0\,\leq\,\varepsilon\,\leq 0.7$, in intervals of $\Delta \varepsilon=0.07$; ii) for a fixed $\varepsilon=0.5$, we varied the trade-off factor in the range 
$0\,\leq\,\beta\,\leq 1.0$, in intervals of $\Delta \varepsilon=0.1$.
The species densities were averaged using $100$ realisations, starting from different initial conditions, in lattices with $500^2$ grid sites, for a timespan of $5000$ generations. The outcomes in Figs.~\ref{fig4a} and \ref{fig4b} show $\rho_i$ regarding the energy depletion and trade-off factors, respectively. Red, dark blue, and cyan stand for species $1$, $2$, and $3$; the yellow line indicates the density of empty spaces. 

Despite individuals of species $1$ performing the trade-off strategy, the only species to benefit is species $2$, whose population rises. According to  Fig.~\ref{fig4a}, species $2$ predominates in the spatial game independently of the depletion energy factor, with $\rho_2$ reaching the maximum for $\varepsilon=0.2$. 
This occurs because, as $\rho_1$ is minimum for $\varepsilon=0.2$, organisms of species $2$ are less threatened. Furthermore, regarding the trade-off factor, as less effort intermediate and low-energy organisms of species $1$ dedicate to reproduction, the more abundant species $2$ is, as shown in Fig.~\ref{fig4b}. For a fixed $\varepsilon=0.5$, the species density $\rho_1$ is minimum for $\beta=0.7$.

We also observe that $\rho_3$ monotonically decreases with $\varepsilon$, as depicted by the cyan line in Fig. \ref{fig4a}. 
This decline is due to the increase in the proportion of intermediate and low-energy organisms of species $i$ as $\varepsilon$ rises, leading to a decrease in the reproduction rate as more individuals allocate energy towards enhancing the dispersal rate. 
Since the number of new organisms of species $1$ drops, the likelihood of individuals in species $3$ failing to obtain energy increases as $\varepsilon$ grows, which increases the 
chances of death due to lack of energy. Likewise, the decline in $\rho_3$ depicted in Fig. \ref{fig4a} can be attributed to the reduction in the natality rate of species $1$ due to the growing prioritisation of mobility over reproduction as $\beta$ increases.

\section{Autocorrelation Function}
Let us now study how individuals of the same species become spatially correlated for varying energy depletion and trade-off factors. 
The autocorrelation function involves finding the inverse Fourier transform of the spectral density, which is calculated by summing the Fourier transform of the spatial distribution of individuals of species each species:
\begin{equation}
C_i(\vec{r}') = \frac{\mathcal{F}^{-1}\{S_i(\vec{k})\}}{C_i(0)},
\end{equation}
where $S(\vec{k})$ is given by
\begin{equation}
S_i(\vec{k}) = \sum_{k_x, k_y}\,\phi_i(\vec{\kappa}),
\end{equation}
and $\phi(\vec{\kappa})$ is the Fourier transform
\begin{equation}
\phi(\vec{\kappa}) = \mathcal{F}\,\{\phi(\vec{r})-\langle\phi\rangle\}.
\end{equation} 
The function $\phi(\vec{r})$ represents the spatial distribution of individuals of species $1$, with $\phi(\vec{r})=0$ and $\phi(\vec{r})=1$ informing if
an individual of species $1$ is present or not at the position $ \vec{r}$ in the lattice, respectively. 

Therefore, the spatial autocorrelation function is given by
\begin{equation}
C_i(r') = \sum_{|\vec{r}'|=x+y} \frac{C_i(\vec{r}')}{min \left[2N-(x+y+1), (x+y+1)\right]}.
\end{equation}
Finally, we define the characteristic length scale, $l_i$, using threshold $C_i(l)=0.15$, with $i=1,2,3$.

Figures \ref{fig5a} and \ref{fig5b} show the dependence of the average length scale
on the energy depletion and trade-off factors, respectively.
The outcomes were obtained by running sets of $100$ realizations varying the random initial conditions in lattices with $500^2$ grid sites over $5000$ generations.
The error bars indicate the standard deviation; the colours follow the scheme in Fig.~\ref{fig1}. 
Our algorithm computes $l_i$ by utilizing the spatial organisms' disposition at the end of each simulation (at time $t=5000$).

The results validate that as $\varepsilon$ and $\beta$ grows,
the scale of the single-species spatial domains decreases. This quantifies the reduction in the aggregate of individuals of the same species observed in the snapshots in Fig.~\ref{fig2}. Irrespective of $\varepsilon$ and $\beta$, organisms of species $1$ - which are the ones to perform the trade-off strategy - are less spatially correlated. According to 
Figs. \ref{fig5a} and \ref{fig5b}, $l_1$ is the shortest characteristic length scale.

Figure \ref{fig5a} indicates that for $\varepsilon \leq 0.35$, $l_2$ and $l_3$ grow if weak organisms of species $1$ assume a trade-off factor of $90\%$. For $\varepsilon > 0.35$, the spatial correlation among conspecifics of every species significantly decreases. Moreover,
Fig. \ref{fig5b} shows that the higher the trade-off factor, the smaller the single-species spatial domains. 

\section{Coexistence Probability}

Finally, we quantify the impact of the reproduction-mobility trade-off strategy on biodiversity in a scenario of changes in environmental conditions.
For this purpose, we compute the coexistence probability as a function of the organisms' mobility for various values of $\varepsilon$ and $\beta$. 
We ran collections of $1000$ simulations in lattices with $100^2$ grid points for $ 0.05\,<\,m\,<\,0.95$ in intervals of $ \Delta\, m\, =\,0.05$. Selection and reproduction probabilities were set to be $s=r\,=\,(1-m)/2$; the timespan is $10000$ generations.
If at least one species disappears before the end, biodiversity is lost. The coexistence probability was determined as the proportion of simulations in which all species were present at the end.

\begin{figure}[t]
\centering
    \begin{subfigure}{.49\textwidth}
        \centering
        \includegraphics[width=85mm]{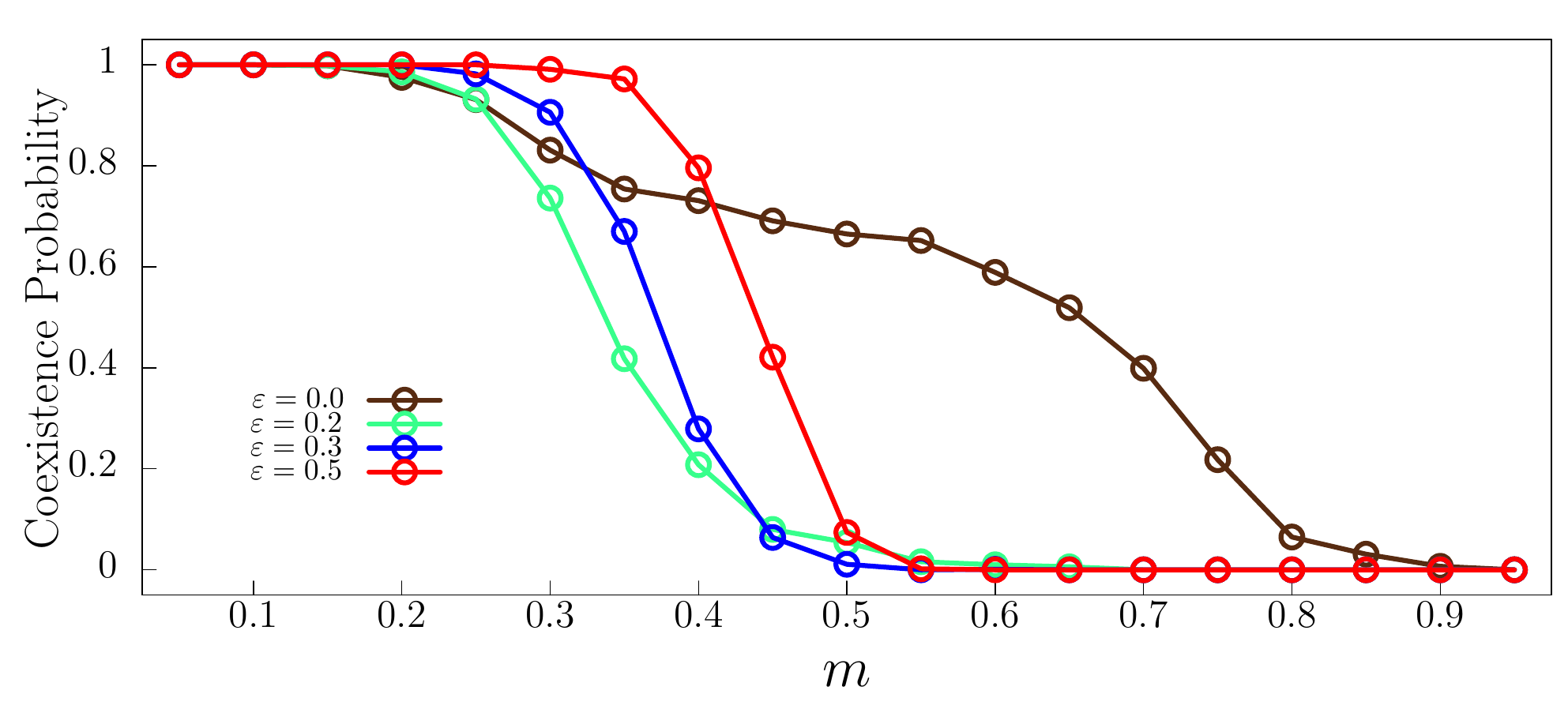}
        \caption{}\label{fig6a}
    \end{subfigure} %
       \begin{subfigure}{.49\textwidth}
        \centering
        \includegraphics[width=85mm]{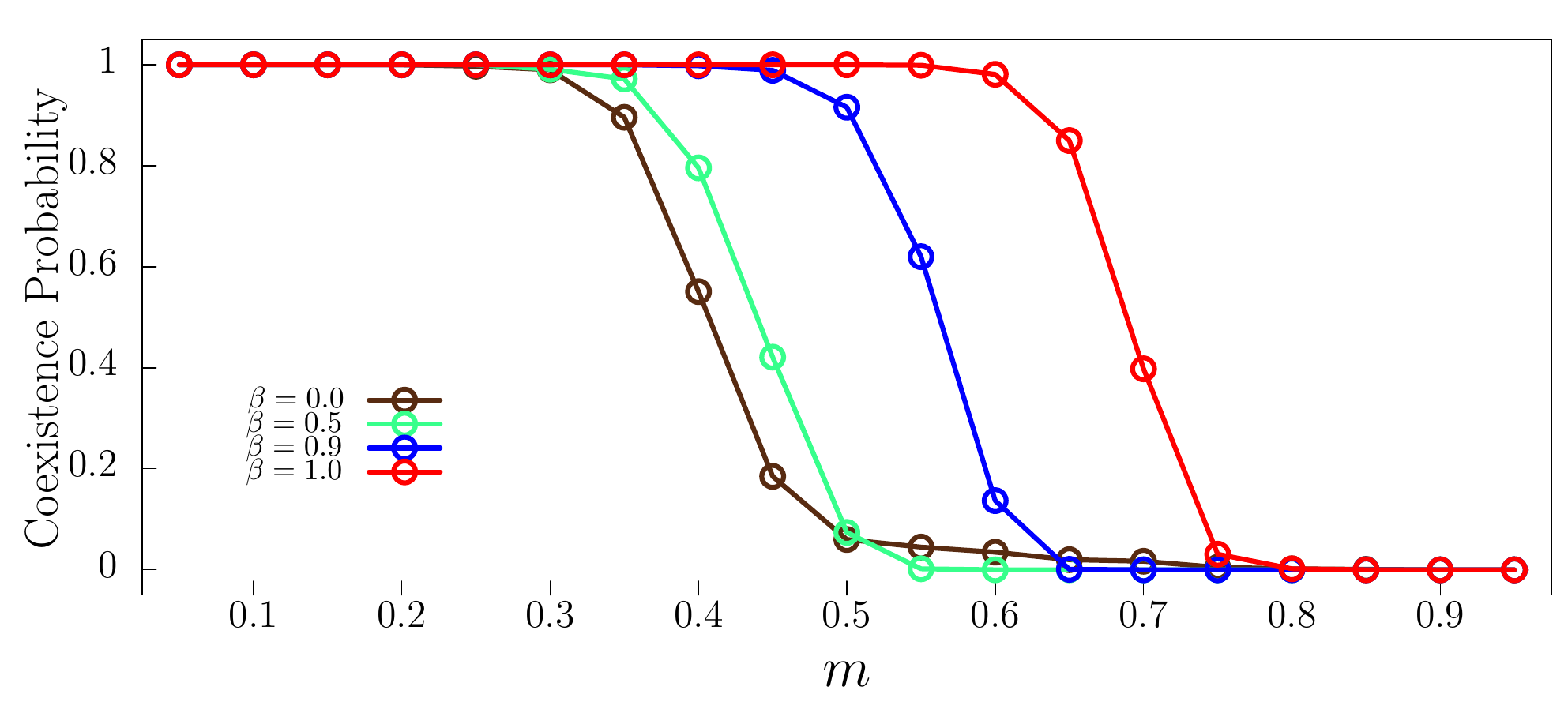}
        \caption{}\label{fig6b}
    \end{subfigure} 
    \caption{Coexistence probability as a function of the mobility probability. In Fig.~\ref{fig6a}, the brown, green, blue, and red lines, depict the outcomes for $\varepsilon=0.0$, $\varepsilon=0.2$, $\varepsilon=0.3$, and $\varepsilon=0.5$, respectively, for a fixed $\beta=0.9$; the same colours are used in Fig.~\ref{fig6b} to show the outcomes for $\beta=0.0$, $\beta=0.5$, $\beta=0.9$, and $\beta=1.0$, respectively, for a fixed $\varepsilon=0.5$. 
The results indicate the proportion of simulations resulting in coexistence in groups of
$1000$ simulations, running in lattices with $100^2$ grid points for a time span of $10000$ generations.}
  \label{fig6}
\end{figure}

As known, 
species diversity is more jeopardised in cyclic systems with high mobility probability. 
According to the brown line in Fig.~\ref{fig6a}, 
not surprisingly, the coexistence probability is higher in the hypothetical scenario where organisms would not suffer energy loss when failing selection interactions, as observed in the brown line of Fig. \ref{fig6a}. 
However, counter-intuitively, the outcomes show that the coexistence probability increases if the energy loss process is accelerated (high $\varepsilon$), as depicted in Fig. \ref{fig6a}, for  $\varepsilon=0.3$ (green line), $\varepsilon=0.5$ (blue line), and $\varepsilon=0.6$ (red line). 

In addition, Fig.~\ref{fig6b} depict the coexistence probability for $\beta=0.0$ (brown line), $\beta=0.5$ (green line), $\beta=0.9$ (blue line), and $\beta=1.0$ (red line). We conclude that the more energy shifts from reproduction to mobility (high $\beta$), the more biodiversity is promoted.
The benefit in the coexistence probability can be attributed to the erosion of spatial patterns caused by the trade-off strategy. This diminishes the average dimensions of single-species spatial domains as the parameter $\beta$ increases, as observed in the spatial patterns in Fig.~\ref{fig2} and quantified in Fig.~\ref{fig5}.
This effect outweighs the reduction in the survival probability of organisms of species $2$ and $3$ caused by the trade-off strategy of individuals of species $1$.

\section{Discussion and Conclusions}

Studying the spatial rock-paper-scissors models, we considered that,
if environmental conditions demand more energy for organisms' survival, energy loss is accelerated whenever an organism of species $i$ fails to eliminate organisms of species $i+1$.
Because of this, our simulations show an erosion 
in the self-organised organisms' spatial organisation leading to a decrease in the characteristic length scale of groups of conspecifics. This is crucial for the perpetuation of species, with biodiversity being further promoted, although the organisms' survival probability.

In addition, another factor contributes to biodiversity promotion: the reduction in the generation of offspring by the species $1$, whose individuals concentrated on increasing its speed.  
We discovered that for a high trade-off factor, individuals of species $1$ 
achieve the goal of increasing survival probability but at the expense of other species whose lifespan is shortened. Again, counterintuitively, the results show that despite the decrease in life expectancy of individuals of species that do not execute the evolutionary reproduction-mobility trade-off, the chances of coexistence increase.
Furthermore, we found that as individuals of species $1$ reproduce less, the number of enemies of species $2$ decreases. Because of this, species $1$ do not prevail in the spatial game, but the predominance is of species $2$.  

Our outcomes can be extended for more complex systems with an odd number of species, where spiral patterns are observed from random initial conditions. Also, our conclusions hold for models with arbitrary energy states. The evolutionary trade-off strategy can significantly impact the average survival time of organisms, with benefits for one species but adverse effects for others.

\section*{Acknowledgments}
We thank CNPq, ECT, Fapern, and IBED for financial and technical support.

\bibliographystyle{elsarticle-num}
\bibliography{ref}

\end{document}